# Multi-agent Path Planning and Network Flow

Jingjin Yu    Steven M. LaValle

*Abstract*— This paper connects multi-agent path planning on graphs (roadmaps) to network flow problems, showing that the former can be reduced to the latter, therefore enabling the application of combinatorial network flow algorithms, as well as general linear program techniques, to multi-agent path planning problems on graphs. Exploiting this connection, we show that when the goals are permutation invariant, the problem always has a feasible solution path set with a longest finish time of no more than $n + V - 1$ steps, in which $n$ is the number of agents and $V$ is the number of vertices of the underlying graph. We then give a complete algorithm that finds such a solution in $O(nVE)$ time, with $E$ being the number of edges of the graph. Taking a further step, we study time and distance optimality of the feasible solutions, show that they have a pairwise Pareto optimal structure, and again provide efficient algorithms for optimizing two of these practical objectives.

## I. INTRODUCTION

Consider the problem illustrated in Fig. 1, which inspired the authors to pursue this research. As an exercise (26-1 in [7]), the *escape problem* is to determine, given $m \leq n^2$ evaders placed on $m$ different points of an $n \times n$ grid, whether there are $m$ vertex disjoint paths from these $m$ locations to $m$ different points on the boundary of the grid. Intended as a demonstration of applications of *maximum flow* algorithms (Ch. 26 of [7]), it undoubtedly mimics multi-agent[1] path planning problems on graphs. Intrigued by the elegant network flow based solution to the escape problem, we wonder: How tightly are these two classes of problems intertwined and how we may take advantage of the relationship?

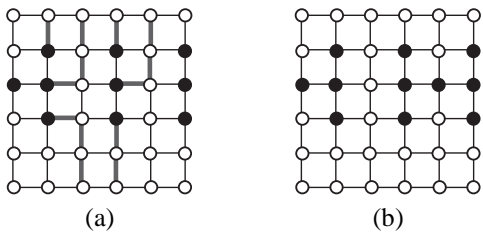

Fig. 1. Examples of the escape problem on a $6 \times 6$ grid. The black discs are the initial evader locations. The goal is to plan disjoint paths for the evaders to reach different vertices on the boundary of the grid. a) An instance with solution given as the bold edges. b) An instance without a solution.

This paper is a preliminary extended version of [?]. This work was supported in part by NSF grants 0904501 (IIS Robotics) and 1035345 (Cyberphysical Systems), DARPA SToMP grant HR0011-05-1-0008, and MURI/ONR grant N00014-09-1-1052.

Jingjin Yu is with the Department of Electrical and Computer Engineering, University of Illinois at Urbana-Champaign, Urbana, IL 61801 USA. E-mail: jyu18@uiuc.edu. Steven M. LaValle is with the Department of Computer Science, University of Illinois at Urbana-Champaign, Urbana, IL 61801 USA. E-mail: lavalle@uiuc.edu.

[1]We use *agent* instead of *robot* since the method applies to scenarios such as evacuation planning.

In this paper, we explore and exploit the connection between multi-agent path planning on *collision-free unit-distance graphs* (or CUGs, see Section II for the definition) and network flow. We begin by showing that multi-agent path planning on CUGs is closely related to a class of problems called *dynamic network flow* or *network flow over time*. We then focus on the permutation invariant multi-agent path planning problem on CUGs (by permutation invariant, we mean that goals are not pre-assigned to agents. Instead, we only require that each goal is reached by a unique agent), establishing that such problems always have solutions. To solve the problem algorithmically, an adapted maximum flow algorithm is provided which plans collision free paths for all agents with worst time complexity $O(nVE)$, in which $n$ is the number of agents, $V$ is the number of vertices of the CUG and $E$ is the number of edges of the CUG. Moreover, we guarantee that the last agent takes time no more than $n + V - 1$ to reach its goal, assuming that agents travel at unit speed. Next, we construct efficient algorithms for obtaining temporally and spatially optimal solutions. For example, our algorithm for shortest overall time has running time $O(nVE \log V)$. We also show that these temporal and spatial objectives cannot be optimized simultaneously (i.e., they have a *Pareto optimal* structure). Portions of the Pareto optimal structure collapse as we move to the permutation-invariant multi-agent path planning problem on CUGs with goal-replacement.

As a universal subroutine in multi-agent systems, collision-free path planning for multiple agents finds applications in tasks spanning assembly [18], [33], evacuation [4], [39], formation control [2], [38], [41], [43], [45], localization [14], object transportation [31], [40], search and rescue [21], and so on. Given its importance, path planning for multi-agent systems has remained as a subject of intense study for many decades. Due to the vast size of the available literature, we only mention a most related subset of the research in this field and refer the readers to [5], [26], [28] and the references therein for a more comprehensive review of the subject.

When all agents are treated as a single agent with a high dimensional configuration space, the problem can be solved using cylindrical algebraic decomposition [6] or Canny's roadmap algorithm [3], in theory. Such *coupled* approaches suffer from the curse of dimensionality; even when sampling based methods [23], [27] are used, instances involving only a small number of agents can be computationally challenging. This difficulty prompts the study of methods that seek to explore local features whenever possible to avoid working with too many agents at a time. Among these, *decoupled* planning is the most popular, which generally performs coor-dination of robot motion after deciding a path for each robot [17], [22], [34], [37], [42], [49]. In contrast, priority based methods force an order on agents to significantly reduce the search space [10], [47]. Some more recent works using decoupling heuristics include applying optimal decoupling techniques to exploit problem instances with low degrees of coupling [48], using *push-and-swap* primitives to avoid unnecessary exploration of search space [30], and heuristics aimed at performance guarantees (completeness is lost) [50].

Our algorithmic efforts in this paper focus on the *permutation invariant* multi-agent path planning problem on CUGs. Such formulations, in both discrete and continuous forms, are extensively studied as formation control problems [2], [38], [41], [43], [45], among others. On research that appears mostly related to this aspect of our paper, a discrete grid abstraction model for formation control was studied in [32]. To plan the paths, a three-step process was used in [32]: 1) Target assignment, 2) Path allocation, 3) Trajectory scheduling. Although it was shown that the process always terminates, no characterization of solution complexity was offered. In contrast, we provide very efficient algorithms that solve a strictly more general class of problems with optimality assurance. On the continuous side, a novel *formation space* approach was employed to represent the entire formation of robot teams with a single polynomial of which the roots correspond to the unassigned configurations for the robots in the formation [24].

We delay the literature review on network flow, from which we devise our time expansion construction for multi-agent path planning, to Section III. The basic idea of applying time expansion to robotics problem is far from new [10], [36]. To the best of our knowledge, however, the research presented here is an original attempt at proposing a general time expansion technique, connecting it to network flow, and making full use of the benefits that come with this approach. We also note that our exact and complete algorithms all come with low constants in their respective worst case time complexity because they are derived from well studied fully combinatorial algorithms[2]. Our simulation result, which we omit due to the length limit, confirms this assertion.

There are three main contributions. First, we formally establish the link between multi-agent path planning on graphs and network flow, showing how multi-agent path planning can be reduced to network flow problems, thereby enabling the potential application of powerful tools from combinatorial optimization to path planning for multiple agents in a principled way. Second, for the planning problem in which agents do not have pre-specified goals, we give fast and complete algorithms for finding collision free path sets that deliver every agent to a different goal. Third, we study time and distance optimality of the feasible solutions to the aforementioned problem, show that they have a pairwise Pareto optimal structure, and again provide efficient algorithms for optimizing two of these practical objectives.

The rest of the paper is organized as follows. In Section II, we define three multi-agent path planning problems on CUGs. Section III starts with a quick review of network flow problems and then proceeds to show the reduction from multi-agent path planning on CUGs to network flow. Concentrating our efforts on the permutation invariant multi-agent path planning problem, Section IV begins with a key construction that allows us to tightly bound the time steps required for a time-expanded network to have a feasible solution, which in turn enables efficient algorithms. Section V takes a further step and studies solution optimality on three natural objectives, showing the objectives have a Pareto optimal structure. Section VI provides a quick discussion of the goal-replacement case, introduces a fourth objective, and shows that the three temporal objectives lose the partial order structure. We conclude in Section VII.

## II. MULTI-AGENT PATH PLANNING PROBLEMS ON COLLISION-FREE UNIT-DISTANCE GRAPHS

Let $G = (V, E)$ be a connected, undirected, simple graph (i.e., no multi-edges), in which $V = \{v_i\}$ is its vertex set and $E = \{(v_i, v_j)\}$ is its edge set. Let $A = \{a_1, \ldots, a_n\}$ be a set of agents with initial and goal locations on $G$ given by the injective maps $x_I : A \to V$ and $x_G : A \to V$, respectively. Note that $A$ is essentially an index set; $x_I(A)$ and $x_G(A)$ are the set of initial and goal locations, respectively. We require that $x_I(A)$ and $x_G(A)$ be disjoint. For convenience, we let $n = |A|$ and use $V, E$ to denote the cardinality of the sets $V, E$, respectively, since the meaning is usually clear from the context. Let $\sigma$ be a bijection[3] that acts on $x_G$, a *feasible path* for a single agent $a_i$ is a map $p_i : \mathbb{Z}^+ \to V$ with the following properties[4]: 1. $p_i(0) = x_I(a_i)$. 2. For each $i$, there exists a smallest $k_{\min} \in \mathbb{Z}^+$ such that $p_i(k_{\min}) = (\sigma \circ x_G)(a_i)$ for some fixed $\sigma$. That is, the end point of the path $p_i$ is some goal vertex. 3. For any $k \geq k_{\min}$, $p_i(k) \equiv (\sigma \circ x_G)(a_i)$. 4. For any $0 \leq k < k_{\min}$, $(p_i(k), p_i(k+1)) \in E$ or $p_i(k) = p_i(k+1)$. Intuitively, think of the domain of the paths as discrete time steps. We say that two paths $p_i, p_j$ are in *collision* if there exists $k \in \mathbb{Z}^+$ such that $p_i(k) = p_j(k)$ (meet) or $(p_i(k), p_i(k+1)) = (p_j(k+1), p_j(k))$ (head-on). If $p(k) = p(k+1)$, the agent stays at vertex $p(k)$ during the time interval $[k, k+1]$.

As mentioned, in this paper, we work with a specific type of graph called the collision-free unit-distance graph (CUG): A CUG is a connected, undirected graph $G$ satisfying the following: 1. Every edge is of unit length; 2. Given any two distinct edges $(u_1, v_1)$ and $(u_2, v_2)$ of $G$ with $u_1 \neq u_2, v_1 \neq v_2$, two disc shapes (or spherical for 3D or more) agents of radius less than $\sqrt{2}/4$ traveling at unit speed through these edges (starting simultaneously at $u_1, u_2$, respectively) will never collide. A radius of $\sqrt{2}/4$ is the largest possible for two adjacent agents to travel along an "L" shaped path. One can easily verify that any graph with unit edge length and no acute angles between adjacent edges is a CUG. Hence, a connected 2D grid with holes is a CUG. Since subgraphs of 2D grids are easy to draw and visualize, we generally use subgraphs of 2D grids when we create examples in this paper. With the above setup, the *multi-agent path planning on CUGs* problem is defined as follows.

---

[2]A *fully combinatorial algorithm* is an algorithm that only adds, subtracts, and compares values; no multiplication and division operations are allowed (i.e., ordered *group* operations versus ordered *field* operations) [16].

[3]$\sigma$ is introduced to unify the problem formulations; its use will become clear shortly. For now, the reader may think of it simply as the identity map.
[4]In this paper, we let $\mathbb{Z}^+ := \mathbb{N} \cup \{0\}$.

**Problem 1** *(Multi-agent Path Planning on CUGs)* Given a 4-tuple $(G,A,x_I,x_G)$ in which $G$ is a CUG, find a set of paths $P = \{p_1,\ldots,p_n\}$ such that $p_i$'s are feasible paths for respective agents $a_i$'s with $\sigma$ being the identity map and no two paths $p_i, p_j$ are in collision.

We require the graph to be a CUG so that it is suitable for multi-agent path planning. We formalize the rationale in the following lemma.

**Observation 2** *Let $p_i, p_j$ be two paths that are not in collision (as a partial solution to Problem 1). Then two disc shaped agents[5] of radius less than $\sqrt{2}/4$, starting at the same time and moving along these respective paths with unit speed, will never collide.*

PROOF. Without loss of generality, assume the two agents (say, $a_i$ and $a_j$) started moving at time $t = 0$ along $p_i$ and $p_j$, respectively. Between any time steps $k$ and $k+1$, $k \geq 0$, agents $a_i$ and $a_j$ travels along edge $(p_i(k), p_i(k+1))$ and $(p_j(k), p_j(k+1))$, respectively. Since $p_i$ and $p_j$ are not in collision, we always have $p_i(k) \neq p_j(k)$, $p_i(k+1) \neq p_j(k+1)$, and $(p_i(k), p_i(k+1)) \neq (p_j(k+1), p_j(k))$. By the definition of CUG, two disc shaped agents of radius less than $\sqrt{2}/4$ traveling on these edges at unit speed will never collide between time interval $[k, k+1]$. Since $k$ is arbitrary, the two agents will never collide. □

Observation 2 shows that a solution to Problem 1 provides a path set for disc agents with radius $\sqrt{2}/4$ in $A$ to reach their respective goals without a collision. It is easy to see that not all instances of this problem are solvable. Moreover, the decision version of Problem 1 (i.e., is there a solution that takes the agents to goals within $K$ time steps) is NP-complete[6]. If we remove the assumption that all agents must reach their respective goals and allow permutation invariant paths (i.e., as long as each goal gets occupied by a unique agent in the end), Problem 1 becomes the *permutation invariant multi-agent path planning on CUGs* problem.

**Problem 3** *(Permutation Invariant Multi-agent Path Planning on CUGs)* Given a 4-tuple $(G,A,x_I,x_G)$ in which $G$ is a CUG, find a set of paths $P = \{p_1,\ldots,p_n\}$ such that $p_i$'s are feasible paths for respective agents $a_i$'s for an arbitrary (but fixed) permutation $\sigma$ and no two paths $p_i, p_j$ are in collision.

---

[5]Or spherical agents with radius less than $\sqrt{2}/4$, for dimensions higher than 2.

[6]The lengthy proof is out of scope for the current paper. For curious readers, the NP-hardness proof is similar to that from [44]; the NP membership proof, which is non trivial, leads to fast heuristics for solving Problem 1.

Problem 3 models the problem in which multiple identical or indistinguishable agents need to be deployed for serving requests at different locations. This problem always has a solution: We simply plan and execute one path at a time and use more "remote" goal vertices earlier to avoid possible blocking of later paths; a proof of the existence of such a choice of paths is given in Section IV. Going one step further and allowing multiple agents to reach the same goal (at different time steps), we get the *permutation invariant multi-agent path planning on CUGs with goal replacement* problem.

**Problem 4** *(Permutation Invariant Multi-agent Path Planning on CUGs with Goal Replacement)* Given a 4-tuple $(G,A,x_I,x_G)$ in which $G$ is a CUG, find a set of paths $P = \{p_1,\ldots,p_n\}$ such that $p_i$'s are feasible paths for respective agents $a_i$'s for an arbitrary (but fixed) $\sigma$ and no two paths $p_i, p_j$ are in collision except at vertices of $x_G(A)$.

This problem provides a realistic model for scenarios such as evacuation of a building due to fire hazard; the set $x_G(A)$ represents the exits to safety: After an agent reaches an exit, the agent can be removed from the system and the exit becomes available again shortly after. Since $G$ is connected and goals can be reused, a feasible solution to Problem 4 can be easily obtained by sending all agents to a single goal vertex sequentially.

### III. MULTI-AGENT PATH PLANNING ON CUGS AND NETWORK FLOW

#### A. Network Flow

In this subsection we give a brief review of network flow problems and algorithms pertinent to our problems. For surveys on network flow, see [1], [13]. We start with the classic static network flow problems.

**Static Network Flow.** A *network* $\mathcal{N} = (G, u, c, S)$ consists of a directed graph $G = (V, E)$ with $u, c : E \to \mathbb{Z}^+$ as the maps defining the capacities and costs on edges, respectively, and $S \subset V$ as the set of sources and sinks. We let $S = S^+ \cup S^-$, with $S^+$ denoting the set of sources and $S^-$ denoting the set of sink vertices. For a vertex $v \in V$, let $\delta^+(v)$ (resp. $\delta^-(v)$) denote the set of edges of $G$ going to (resp. leaving) $v$. A feasible (static) $S^+, S^-$-flow on this network $\mathcal{N}$ is a map $f : E \to \mathbb{Z}^+$ that satisfies edge capacity constraints,

$$\forall e \in E, \quad f(e) \leq u(e), \quad (1)$$

the flow conservation constraints at non terminal vertices,

$$\forall v \in V \backslash S, \sum_{e \in \delta^+(v)} f(e) - \sum_{e \in \delta^-(v)} f(e) = 0, \quad (2)$$

and the flow conservation constraints at terminal vertices,

$$\sum_{v \in S^+}(\sum_{e \in \delta^-(v)} f(e) - \sum_{e \in \delta^+(v)} f(e)) = \\ \sum_{v \in S^-}(\sum_{e \in \delta^+(v)} f(e) - \sum_{e \in \delta^-(v)} f(e)). \quad (3)$$

The quantity on either side of (3) is called the *value* of the flow.

The classic (single-commodity) *maximum flow* problem asks the question that given a network $\mathcal{N}$, what is the maximum value of flow that can be pushed through the network (i.e., seeking to maximize $F$)? The *minimum cost maximum flow* problem further requires the flow to have minimum total cost among all maximum flows. That is, we want to find the flow among all maximum flows such that the quantity

$$\sum_{e \in E} c(e) \cdot f(e) \quad (4)$$

is minimized. Given integer inputs, integer maximum flow always exists, and many polynomial time algorithms exist for finding such a solution [9], [15]. The minimum cost maximum flow problem is equivalent to the *minimum cost circulation* problem, which is also solvable in polynomial time [46].

When additional structure is put on $S$, additional questions arise. If we limit the supply (demand) of the source (sink) vertices, we obtain a type of the flow problem called the *transshipment* problem. To formalize this, let $d: V \to \mathbb{Z}$ be the supplies on the vertices of $G$. Given a vertex $v \in V$, a positive $d(v)$ suggests that the vertex has positive supply ($v \in S^+$) and a negative one suggests that the vertex has positive demand ($v \in S^-$). For all other vertices $v$, $d(v) = 0$. The basic version of the transshipment problem asks for a feasible flow through the network that also respects the supply/demand requirements

$$\begin{aligned} \forall v \in S^+, \quad &\sum_{e \in \delta^-(v)} f(e) - \sum_{e \in \delta^+(v)} f(e) = d(v), \\ \forall v \in S^-, \quad &\sum_{e \in \delta^+(v)} f(e) - \sum_{e \in \delta^-(v)} f(e) = d(v). \end{aligned} \quad (5)$$

The transshipment problem becomes the *evacuation* problem when $|S^-| = 1$ and the demand of the single sink vertex is equal to the total supply of the source vertices. The transshipment problem and the evacuation problem, as special cases of the maximum flow problem, can be solved with maximum flow algorithms mentioned above. If we instead require that vertices of $S^+, S^-$ are paired up as $(s_1, s_1'), \ldots, (s_k, s_k')$ and that commodity of type $i$ can be injected only into $s_i$ and taken out at $s_i'$, we get the *multi-commodity flow* problem. Optimality questions as these from the single-commodity case can be asked here as well. Unlike in the single commodity case, finding integer maximum flow for multi-commodity problems is NP-hard in general and MAX SNP-hard (NP-hard to approximate below a certain multiple of optimal flow value) even for some simple restrictions [8].

**Dynamic Network Flow**. If we consider that flowing commodities through edges takes some time to complete, the problem becomes a *dynamic network flow* problem, which sometimes is also called *network flow over time*. There are two common variations of the dynamic network flow model:

*Discrete time* and *continuous time*. In a *discrete time* model, flows enter and exit from vertices at integer time steps $t = 0, 1, \ldots, T$. For a given edge $e = (u, v) \in E$, we may view the cost $c(e)$ as the time that is required to pass an amount of flow (not exceeding the capacity) from the tail $u$ to the head $v$ of the edge $e$. Therefore, we may interpret a (static) flow network $\mathcal{N}$ as a dynamic one without any change of notations. In the closely related *continuous time* model, which we do not use in this paper, a *flow rate* is assigned to each edge, designating how fast a unit of flow can pass through the edge. The constraints imposed in the static network flow model generally apply to dynamic network flow models, except that dynamic network flow further requires that at any time, the flow passing through any edge cannot exceed the edge capacity.

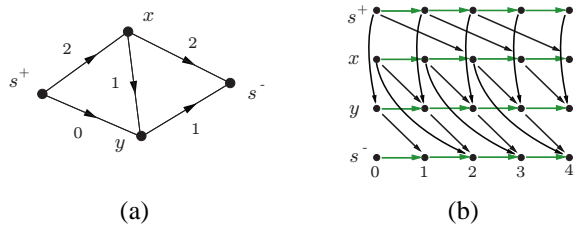

Fig. 2. a) The (static) flow network with source $s^+$ and sink $s^-$. The numbers on the edges are the costs/time delay for passing through these edges. We may assume that the capacities are all unit capacities. b) The time-expanded network with 5 copies of the original vertices ($T = 4$). All edges have unit capacity. There is a forward edge between two vertices $u$ and $v$ at time steps $t$ and $t'$, respectively (e.g. $x$ at $t = 0$ and $y$ at $t' = 1$), if one of the following is true: 1. $e = (u, v)$ is an edge of the static network with $c(e) = t' - t$ (the black edges, which retain the costs as $c(e)$'s); 2. $u, v$ are the same vertex of the static network and $t' - t = 1$ (the green edges, which have unit costs). The green edges are also called *holdover* edges since traveling through a green edge is the same as the agent not actually moving.

Given a dynamic flow network, a question similar to the single-commodity maximum flow problem is the following: Starting at $t = 0$, what is the maximum units of flow the can reach the sinks on or before time $t = T$? It turns out that this problem can be solved using static flow algorithms such as Edmonds-Karp [9] over a *time-expanded network*. For example, given the dynamic flow network in Fig. 2(a), its time-expanded network with $T = 4$ is given in Fig. 2(b). To compute a flow over the time expanded network, we first add a *super source* and connect it using outgoing edges to all copies of source vertices at $t = 0$, and add a *super sink* and connect all copies of sink vertices for all $t$ to it using outgoing edges (the super source, super sink, and additional edges are not shown in Fig. 2(b)).

**Lemma 5** *For a sufficiently large $T$, a flow for a dynamic flow network $\mathcal{N}$ is feasible if and only if the corresponding static flow on the time-expanded network of $\mathcal{N}$ is feasible.*

A proof of Lemma 5 can be found in [12]. Note that determining a minimally sufficient $T$ required by Lemma 5, which directly affects the running time of the resulting algorithm, is non-trivial. The standard maximum flow algorithms have time complexity depending polynomially on $T$ and are therefore *pseudopolynomial* in general. For a special class of problems, the *quickest transshipment problem*, of which the goal is finding the quickest feasible flow for a transshipment problem over a dynamic network, strongly polynomial time algorithm[7] exists [19]. However, the algorithm requires calling subroutines (for example, submodular function optimization routines) that are not *fully combinatorial* algorithms and also has with large constant terms when it comes to asymptotic time complexity. As we will see, our problems can be solved using polynomial time fully combinatorial algorithms, due to their special structures.

*B. Equivalence Between Multi-agent Path Planning on CUGs and Maximum Network Flow*

In this subsection, we establish a reduction from the problems of our interest to multi-commodity network flow. For illustration purpose, we use the simple graph $G$ in Fig. 3(a), with initial locations $\{s_i^+\}, i = 1,2$ and goal locations $\{s_i^-\}, i = 1,2$. An instance of Problem 1 is given by $(G, \{a_1, a_2\}, x_I : a_i \mapsto s_i^+, x_G : a_i \mapsto s_i^-)$. To be able to apply maximum flow algorithms, we construct from $G$ a time-expanded directed graph $G'$, part of which is shown in Fig. 3(c). We construct Fig. 3(c) as follows.

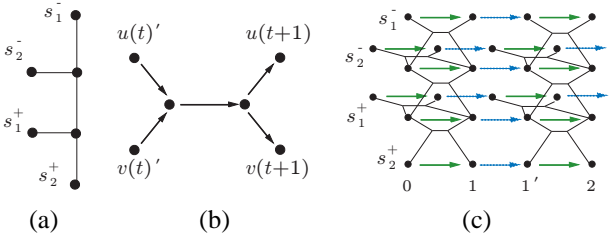

Fig. 3. a) A simple CUG $G$. b) A gadget for splitting an undirected edge through time steps. c) Part of the time-expanded network ($T = 2$).

Since we cannot create an infinite time-expanded network, we need to specify the required number of time steps. For now assume that this number is some sufficiently large $T$ (that is, if a flow with value $F$ is achievable with an arbitrarily long time expansion, then $F$ is also achievable with only $T$ time steps). After fixing $T$, we create $2T + 1$ copies of vertices from $G$, with indices $0, 1, 1', \ldots$, as shown in Fig. 3(c). For each vertex $v \in G$, we denote these copies $v(0) = v(0)', v(1), v(1)', v(2), \ldots, v(T)'$. For each edge $(u, v) \in G$ and time steps $t, t+1$, $0 \leq t < T$, we then add the gadget shown in Fig. 3(b) between $u(t)', v(t)'$ and $u(t+1), v(t+1)$ (arrows from the gadget are omitted from Fig. 3(c) since they are too small to draw). This gadget ensures that two agents cannot travel in opposite directions on an edge in the same time step. For the gadget, we assign unit capacity to all edges, unit cost to the horizontal middle edge, and zero cost to the other four edges. To finish the construction of Fig. 3(c), for each vertex $v \in G$, we add one edge between every two successive copies (i.e., we add the edges $(v(0), v(1)), (v(1), v(1)'), \ldots, (v(T), v(T)'))$. These correspond to the green and blue edges in Fig. 3(c). For all green edges, we assign them unit capacity and cost; for all blue edges, we assign them unit capacity and zero cost.

The graph Fig. 3(c) is the main piece of $G'$, which is mostly done with the exception of the set $S$. We may simply let $S^+ = \{u(0) : u \in \{s_i^+\}\}$ and $S^- = \{v(T)' : v \in \{s_i^-\}\}$. That is, $S^+$ contains the first copies of the initial locations and $S^-$ the last copies of the goal locations. The network $\mathcal{N}' = (G', u, c, S^+ \cup S^-)$ is now complete; we have reduced Problem 1 to an integer maximum multi-commodity flow problem on $\mathcal{N}'$ with each agent from $A$ as a single type of commodity.

**Theorem 6** *Given an instance of Problem 1 with input parameters $(G, A, x_I, x_G)$, there is a bijection between its solutions (with maximum number of time steps up to $T$) and the integer maximum multi-commodity flow solutions of flow value $n$ on the time-expanded network $\mathcal{N}'$ constructed from $(G, A, x_I, x_G)$ with $T$ time steps.*

PROOF.

(Injectivity) Assume that $P = \{p_1, \ldots, p_n\}$ is a solution to an instance of Problem 1. For each $p_i$ and every time step $t = 0, \ldots, T$, we mark the copy of $p_i(t)$ and $p_i(t)'$ (recall that $p_i(t)$ corresponds to a vertex of $G$) at time step $t$ in the time-expanded graph $G'$. Connecting these vertices of $G'$ sequentially (there is only one way to do this) yields one unit of flow $f_i$ on $\mathcal{N}'$ (after connecting to appropriate source and sink vertices in $S^+, S^-$, which is trivial). It is straightforward to see that if two paths $p_i, p_j$ are not in collision, then the corresponding flows $f_i, f_j$ on $\mathcal{N}'$ are vertex disjoint paths and therefore do not violate any flow constraint. Since any two paths in $P$ are not in collision, the corresponding set of flows $\{f_1, \ldots, f_n\}$ is feasible and maximal on $\mathcal{N}'$.

(Surjectivity) Assume that $\{f_1, \ldots, f_n\}$ is a integer maximum multi-commodity flow on the network $\mathcal{N}'$ with $|f_i| = 1$. First we establish that any pair of flows $f_i, f_j$ are vertex disjoint. To see this, we note that $f_i, f_j$ (both are unit flows) cannot share the same source or sink vertices due to the unit capacity structure of $\mathcal{N}'$ enforced by the blue edges. If $f_i, f_j$ share some non-sink vertex $v$ at time step $t > 0$, both flows then must pass through the same blue edge (see Fig. 3(b)) with $v$ being either the head or tail vertex, which is not possible. Thus, $f_i, f_j$ are vertex disjoint on $\mathcal{N}'$. We can readily convert each flow $f_i$ to a corresponding path $p_i$ (after deleting extra source vertex, sink vertices, vertices in the middle of the gadgets, and tail vertices of blue edges) with the guarantee that no $p_i, p_j$ will collide due to "meet". By construction of $\mathcal{N}'$, the gadget we used ensures that "head-on" collision is also not possible. The set $\{p_1, \ldots, p_n\}$ is then a solution to Problem 1. □

Since integer maximum multi-commodity flow is NP-hard, the above construction does not directly offer an efficient solution to Problem 1. Nevertheless, with backtracking, it is not hard to design complete algorithms that search the time-expanded network $\mathcal{N}'$ for a feasible solution. Our preliminary analysis shows that when the problem instance is not particularly hard (i.e., when there are not many narrow passages in the combined configuration space), a solution can be found relatively quickly. We plan to study this problem in more detail in future work. Alternatively, we may readily obtain an integer linear programming problem from $\mathcal{N}'$ and apply heuristics such as *branch and bound* method [25], for which many heavily optimized numerical packages are readily available. Moving to Problem 3, allowing an arbitrary permutation $\sigma$ to act on $x_G$ means that we may treat all agents as a single type of commodity. Theorem 6 gives us the following corollary.

---

[7]An algorithm is a strongly polynomial algorithm if: 1. The number of operations in the arithmetic model of computation is bounded by a polynomial in the number of integers in the input instance, and 2. The space used by the algorithm is bounded by a polynomial in the size of the input [16].

**Corollary 7** *Given an instance of Problem 3 with input parameters $(G,A,x_I,x_G)$, there is a bijection between its solutions (with maximum number of time steps up to $T$) and the integer maximum flow solutions of flow value $n$ on the time-expanded network $\mathcal{N}'$ constructed from $(G,A,x_I,x_G)$ with $T$ time steps.*

At this point, Problem 3 can be solved in polynomial time using the algorithm for the quickest transshipment problem with linear programming subroutines for optimizing submodular functions. In the next section, we show that we can do better by bounding the required time steps for finding a feasible solution to Problem 3 and then apply more standard combinatorial algorithms for network flow.

## IV. EFFICIENT COMBINATORIAL ALGORITHMS FOR PERMUTATION INVARIANT MULTI-AGENT PATH PLANNING ON CUGS

If we choose to apply combinatorial network flow algorithms over the time-expanded network to find solutions to Problem 3, the first priority is to determine the required number of time steps necessary to find a solution; otherwise we cannot declare that the algorithm is complete. We now provide a tight bound on $T$. Let $(G,A,x_I,x_G)$ be an instance of Problem 3. We first prove some intermediate results on path sets over $G$. To distinguish these paths from the solution path set, denote them as $Q = \{q_1,\ldots,q_n\}$. For convenience, $head(q_i)$, $tail(q_i)$, and $len(q_i)$ denote the start vertex, end vertex, and length of $q_i$, respectively. With a slight abuse of notation, $V(\cdot), E(\cdot)$ denote the vertex set and edge set of the input parameter, which can be either a path, $q_i$, or a set of paths, such as $Q$. An *intersection* between two paths is a maximal consecutive sequence of vertices and edges common to the two paths. A *standalone* goal vertex is a vertex $v \in x_G(A)$ such that there is a single path $q \in Q$ containing $v$. To start off, we want a path set $Q$ with the following properties:

**Property 8** *For all $1 \leq i \leq n$, $head(q_i) \in x_I(A)$ and $tail(q_i) \in x_G(A)$. For any two paths $q_i, q_j$, $head(q_i) \neq head(q_j)$ and $tail(q_i) \neq tail(q_j)$.*

**Property 9** *Each path $q_i$ is a shortest path between $head(q_i)$ and $tail(q_i)$ on $G$.*

**Property 10** *The total length of the path set $Q$ is minimal.*

**Property 11** *If we orient the edges of every path $q_i \in Q$ from $head(q_i)$ to $tail(q_i)$, no two paths share a common edge oriented in different directions.*

Properties 8 and 9 are merely restrictions to have the initial and goal vertices paired up using shortest paths. Property 10 requires the total length of these paths to be minimal. Property 11, which is implied by Property 10, lends to show that the paths can be oriented to form a *directed acyclic graph*.

**Lemma 12** *There exists a set of paths $Q = \{q_1,\ldots,q_n\}$ that satisfies Properties 8-11.*

PROOF.
The first property is trivial to satisfy: Given an arbitrary instance of Problem 3, an arbitrary pairing of vertices, one from $x_I(A)$ and one from $x_G(A)$, will meet the requirement. Since $G$ is connected, for each pair of vertices of the form $(s_i^+, s_i^-)$ with $s_i^+ \in x_I(A)$ and $s_i^- \in x_G(A)$, there is a path between them with finite distance. Picking an arbitrary shortest path $q_i$ between $s_i^+, s_i^-$ on $G$ (there may be multiples of these; for example when $G$ is a grid graph) for $1 \leq i \leq n$ satisfies the second property. The third requirement can be satisfied by checking all possible path sets satisfying the first two requirements and select one with the smallest total distance.

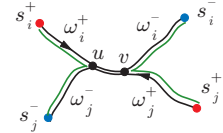

Fig. 4. Two opposite running paths, $q_i = s_i^+ \omega_i^+ uv\omega_i^- s_i^-$ and $q_j = s_j^+ \omega_j^+ vu\omega_j^- s_j^-$ (black paths), have total length at least 2 longer than that of $q_i' = s_i^+ \omega_i^+ \omega_j^- s_j^-$ and $q_j' = s_j^+ \omega_j^+ \omega_i^- s_i^-$ (green paths).

With Properties 8-10 satisfied, we claim that Property 11 is automatic. Suppose the statement is false and assume that two oriented paths $q_i, q_j$ run in different directions on some common edge $(u,v)$. We may write the paths as $q_i = s_i^+ \omega_i^+ uv\omega_i^- s_i^-$ and $q_j = s_j^+ \omega_j^+ vu\omega_j^- s_j^-$, in which $\omega_i^+$ is the path of $q_i$ connecting $s_i^+$ to $u$ (see Fig. 4). $\omega_i^-, \omega_j^+, \omega_j^-$ are interpreted similarly. Then, the paths $q_i' = s_i^+ \omega_i^+ \omega_j^- s_j^-$ and $q_j' = s_j^+ \omega_j^+ \omega_i^- s_i^-$ have total length equaling $len(q_i) + len(q_j) - 2$, which contradicts the shortest total distance assumption. We conclude that no two oriented paths can have edges oriented in opposite directions. □

Above proof technique can be generalized to show that the oriented paths cannot form any directed cycles.

**Theorem 13** *A path set $Q$ that satisfies Properties 8-10 induces a directed acyclic graph (DAG) structure on $E(Q)$.*

PROOF.
Since Property 11 is implied by Properties 8-10, all edges of $E(Q)$ have a unique orientation. Therefore, the path set $Q$ induces a directed graph structure over $E(Q)$. This implies that the statement of the theorem can only be false if there is a directed cycle in the induced directed graph. Since a single path from $Q$, being a shortest path, cannot form a directed cycle itself, at least two or more paths, say $q_1,\ldots,q_k$, are need to form a directed cycle. Without loss of generality, we assume these $k$ paths are all needed to form a cycle (removing any $q_i$, $1 \le i \le k$ will leave no directed cycles by $\{q_1,\ldots,q_k\}\setminus q_i$). That is, for each $1 \le i \le k$, the directed cycle, say $C$, has at least one edge that belongs only to $q_i$ (an illustration is given in Fig. 5). We show that we can update these paths, without violating Properties 8-10, to obtain a path in the end that intersects itself, which contradicts Property 10.

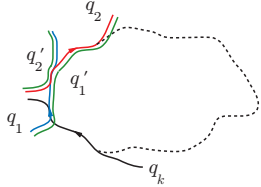

Fig. 5. A hypothetical (directed) cycle in a path set. We can switch the heads and tails of $q_1, q_2$ to get the green paths $q_1', q_2'$ without changing the total length. Now we can remove $q_2'$ and still have the same directed cycle. Performing the same procedure (essentially an inductive argument) eventually yields a path that contains a directed cycle.

We may write $q_1$ as $\omega_1 u \omega_2 v \omega_3$, in which $u\omega_2 v$ is the maximal segment of $q_1$ that belongs to the cycle $C$; $\omega_1, \omega_3$ may be empty. Some other path intersecting $C$ must intersect $u\omega_2 v$ at $v$ (by the maximality of $u\omega_2 v$) and have a segment belonging to $C$ starting at vertex $v$; let $q_2$ be such a path. Since $q_2$ contributes some unique edges to $C$, there are some edges of $q_2$ in $C$ that follow $v$ but do not belong to $u\omega_2 v$. We can then write $q_2 = \omega_4 v \omega_5 w \omega_6$, in which $w$ is the last vertex of $q_2$ belonging to $C$. Note that $u\omega_2 v$ and $\omega_4 v$ may have edges that overlap. We can rearrange $q_1, q_2$ into $q_1' = \omega_1 u \omega_2 v \omega_5 w \omega_6$ and $q_2' = \omega_4 v \omega_3$. Clearly, $len(q_1) + len(q_2) = len(q_1') + len(q_2')$; that is, the new path set again satisfies Properties 8-10. We have shown that a path set $\{q_1,\ldots,q_k\}$ with a directed cycle can be rearranged to yield a path set such that $\{q_1', q_3,\ldots,q_k\}$ again contains the same directed cycle. Applying the same reasoning recursively shows that we can obtain a shortest path that intersects itself, which is not possible. □

Theorem 13 implies the following.

**Corollary 14** *A path set $Q$ that satisfies Properties 8-10 has a standalone goal vertex.*

PROOF.
By Theorem 13, the paths in $Q$ induces a directed acyclic graph on $E(Q)$. This implies that at least one vertex from $x_G(A)$ must be a standalone goal vertex; otherwise every goal must be on another path and the directed path containing the goals must close to form a directed cycle because there are only a finite number of goals. □

The existence of a standalone goal vertex allows the construction of a path set which decomposes into paths that can be sequentially scheduled without colliding into each other. We characterize such a path set as one with an additional property.

**Lemma 15** *There exists a path set $Q$ satisfying Properties 8-11 and the following additional property:*

**Property 16** *Let $Q_i := \{q_i,\ldots,q_n\}$. For any $1 \le i \le n$, restricting to $Q_i$, among all possible paths connecting an initial location (of $Q_i$) to a standalone goal location (of $Q_i$) using oriented edges from $E(Q_i)$, $q_i$ is one shortest such.*

PROOF.
We begin with a $Q = \{q_1,\ldots,q_n\}$ satisfying Properties 8-10 and construct a $Q' = \{q_1',\ldots,q_n'\}$ satisfying the desired property while preserving Properties 8-10. By Corollary 14, there are one or more standalone goal vertices. Among all possible paths connecting some initial and standalone goals using oriented edges from $E(Q)$, we pick one among the shortest. This is $q_1'$. Note it is likely that $q_1' \notin Q$, in which case we may assume $head(q_1') = head(q_i)$ and $tail(q_1') = tail(q_j)$ for some $q_i, q_j \in Q$. There are two possibilities: Either $E(q_1') \subset E(q_i) \cup E(q_j)$ or $q_1'$ contains edges from some other paths. For the first case (Fig. 6(a)), rearranging the paths as shown in green does not change total path length. I.e., Properties 8-10 still hold. For the second case, we may assume that $E(q_1') \setminus (E(q_i) \cup E(q_j))$ belong to some other paths $q_k$ (applying similar reasoning used in the first case, we can always get such a $q_k$ via switching heads and tails of paths without changing the total path length). The switching shown in Fig. 6(b) gives us $q_1'$ without changing total path length. After updating $Q$ (now contains $q_1'$ as an element), we apply the same procedure to $Q\setminus\{q_1'\}$ and so on; the end result is a path set satisfying all desired properties. □

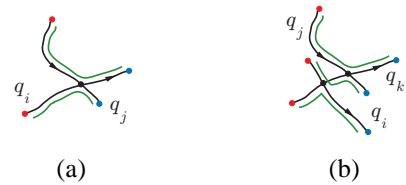

Fig. 6. Possible cases for rearranging paths without affecting total length. a) $q_1'$ is the lower green path. b) $q_1'$ is the middle green path.

If we schedule agents using a path set $Q$ satisfying properties 8-16, there can never be cases where two agents block each other, as a direct consequence of Lemma 12. There is still the possibility that one agent blocks another. The following theorem shows that such blocking can be minimized.

**Theorem 17** *Given an instance of Problem 3 with input parameters $(G,A,x_I,x_G)$ and let $\ell$ be the largest pairwise distance between a member of $x_I(A)$ and a member of $x_G(A)$,*

$$\ell = \max_{\forall u \in x_I(A), v \in x_G(A)} dist(u,v). \qquad (6)$$

*A time-expanded network $\mathcal{N}'$ with $T = n+\ell-1$ is necessary and sufficient for a feasible solution to Problem 3 to exist.*

PROOF.

(Necessity) We construct an instance of Problem 3 shown in Fig. 7. The graph $G$ is two stars with centers connected by a single path; the red vertices form $x_I(A)$ and the blue ones $x_G(A)$. It is clear that all red vertices are of distance $\ell$ to all blue vertices. Given this graph $G$, only one agent can go from a red vertex to the adjacent black vertex at a single time step. As such, it takes at least $n$ time steps for the last agent at a red vertex to go to the neighboring black vertex. After that, it takes the last agent $\ell-1$ steps to reach a blue vertex. Therefore, a total of $T = n+\ell-1$ time steps is necessary.

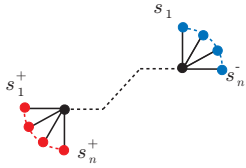

Fig. 7. An instance of Problem 3 for demonstrating necessity claim of Theorem 17.

(Sufficiency) Lemma 15 guarantees the existence of a path set satisfying Properties 8-16; we work with such a path set $Q$ and schedule it to get a path set $P$ in the time-expanded network. The schedule is fairly simple: At each $t = i-1, 1 \leq i \leq n$, we let agent $a_i$ move along $q_i$. The claim is that no collision will occur withing the scheduled path set $P$, which we prove via induction. For the base case, $a_1$ starts at $head(q_1)$ at $t = 0$. By Property 16, no other initial locations can be closer to $tail(q_1)$ than $head(q_1)$. Since other paths start later, they cannot get in the way of $q_1$'s schedule, which we denote $p_1$. Therefore, $p_1$ cannot collide with any other scheduled paths before it reaches its goal.

For the inductive case, assume that $\{q_1, \ldots, q_{k-1}\}$ can be scheduled to get $\{p_1, \ldots, p_{k-1}\}$ without collision. We need to show that $\{q_1, \ldots, q_k\}$ can be scheduled to get $\{p_1, \ldots, p_k\}$ without collision. We use the property that $tail(q_1)$ is a standalone goal vertex, which means that no other $q_i, i \neq 1$ will ever pass $tail(q_1)$. That is, $p_1$ cannot collide with any other path on or after the time it reaches its goal, $tail(q_1)$. We can effectively remove $q_1$ from the set $\{q_1, \ldots, q_k\}$ and apply induction hypothesis to $\{q_2, \ldots, q_k\}$ to see that $\{p_2, \ldots, p_k\}$ contains no pairs that will collide. Adding $p_1$ back proves the inductive case.

Since $len(q_i) \leq \ell$ and no schedule delays a path by more than $n-1$ time steps, $T = n+\ell-1$ is sufficient for schedule the path set $Q$. □

Since $\ell$ cannot be larger than $V$, the number of vertices of $G$, the following corollary is immediate.

**Corollary 18** *For every instance of Problem 3, a feasible solution exists.*

In particular, the construction in the proof of Lemma 12 yields a complete (albeit not necessarily efficient) algorithm for Problem 3. In addition to confirming that any maximum flow algorithm over the time expanded network $\mathcal{N}'$ with $T = n+\ell-1$ is a also complete algorithm, Theorem 17 enables us to show that such algorithms are efficient. A *combinatorial algorithm* is an algorithm that only adds, subtracts, and compares values; no multiplications and divisions are allowed (i.e., ordered *group* operations versus *field* operations). An algorithm is a strongly polynomial algorithm if: 1. The number of operations in the arithmetic model of computation is bounded by a polynomial in the number of integers in the input instance, and 2. The space used by the algorithm is bounded by a polynomial in the size of the input [16].

**Theorem 19** *Problem 3 is solvable using a combinatorial algorithm in strongly polynomial time.*

PROOF.

Since $\ell$ is the length of some shortest path on a connected graph $G$, $\ell = V - 1$ in the worst case. Hence, in the worst case, $T = n + V - 2$. Following the construction algorithm given in Section III, the time-expanded network $\mathcal{N}'$ has an underlying directed graph $G'$ with $O(T)$ copies of the undirected graph $G$ (the gadget and extra edges only introduce a small constant). The graph $G'$ then has $O(V(n+V-2))$ vertices and $O(E(n+V-2))$ edges. Denoting the running time of an maximum flow algorithm as $\text{MF}(n, n_v, n_e)$ in which $n, n_v, n_e$ are the number of agents, vertices, and edges, respectively, Problem 3 can be solved in time $O(\text{MF}(n, V(n+V-2), E(n+V-2))) \sim O(\text{MF}(n, V^2, VE))$ (there cannot be more agents than vertices), which is strongly polynomial since many polynomial time maximum flow algorithms exist. □

Using the Ford-Fulkerson algorithm [11], the time complexity is $O(nVE)$. In practice, even better running times are possible. If $G$ is a planar graph, we have $E \sim O(V)$ and $\ell \sim O(V^{\frac{1}{2}})$. The time complexity then becomes $O(\text{MF}(n, V(n+V^{\frac{1}{2}}-2), V(n+V^{\frac{1}{2}}-2))) \sim O(\text{MF}(n, V(n+V^{\frac{1}{2}}), V(n+V^{\frac{1}{2}})))$. Since in our case $n < V$, Ford-Fulkerson gives us the running time $O(nV(n+V^{\frac{1}{2}})) = O(n^2V + nV^{\frac{3}{2}})$.

## V. OPTIMAL SOLUTIONS

In this section, we present optimal solutions for the permutation invariant multi-agent path planning problem. After introducing several temporal and spatial objectives of practical importance, we apply techniques from network flow to obtain optimal solutions for two of these objectives. Since these objectives are different from the basic version of Problem 3, we provide updated bounds on $T$, the number of times steps sufficient for a solution to exist. Lastly, we show that these objectives possess a Pareto optimal structure and they cannot be optimized simultaneously.

### A. Optimizing over the Feasible Solutions

Having found *feasible* solutions to Problem 3, we turn the focus to the *optimality* of these solutions for practical purposes. As mentioned in Section II, we intend to use the formulation as a model for scenarios such as multi-robot servicing. For many applications, time optimality is a top priority. Optimizing over the feasible solutions to Problem 3 (that is, we require that all goals are reached), there are two natural criteria for measuring time optimality:

**Objective 20** *Minimizing the average time it takes all agents to reach their goals.*

**Objective 21** *Minimizing the time it takes for the last agent to reach its goal.*

In terms of agents (robots or people) serving requests, Objective 20 seeks to minimize the average time before a request gets served, which is arguably the most efficient approach. The time steps sufficient for an optimal solution to exist for this objective is $(n-1)(n-2)/2 + V$.

**Theorem 22** *There exists an optimal solution for Objective 20 in a time-expanded network with $T = (n-1)(n-2)/2 + V$.*

PROOF. Let $P = \{p_1, \ldots, p_n\}$ be an solution path set with minimum total arrival time. The total completion time for $P$ cannot be less than $\sum_{i=1}^{n} len(p_i)$, since that is the shortest possible distance to travel (more time may be needed since a robot may stay idle at an intermediate vertex). If we leave any path $p_k$ out, the total arrival time it takes for all other $p_i$'s to finish cannot be less than $\sum_{i=1}^{n} len(p_i) - len(p_k)$. For any $p_k$, $len(p_k) \leq V - 1$, because interchangeability of robots means that it is never necessary for any robot to revisit a vertex. On the other hand, if we start with a shortest unscheduled path set $Q = \{q_1, \ldots, q_n\}$ and schedule them according to Schedule **??**, the total finish time cannot be more than

$$\sum_{i=1}^{n}(len(q_i) + i - 1) = \sum_{i=1}^{n} len(q_i) + \frac{n(n-1)}{2}. \quad (7)$$

This provides an upper bound on the minimum total arrival time. By the minimality of path set $Q$, we have (the fact that $V - 1 - len(p_k) \geq 0$ is used)

$$\sum_{i=1}^{n} len(p_i) - len(p_k) + V - 1 \geq \sum_{i=1}^{n} len(q_i). \quad (8)$$

Since the total time it takes for paths other than $p_k$ to finish is at least $\sum_{i=1}^{n} len(p_i) - len(p_k)$, the time it takes for any path $p_k \in P$ to finish cannot be longer than

$$\sum_{i=1}^{n} len(q_i) + \frac{n(n-1)}{2} - (\sum_{i=1}^{n} len(p_i) - len(p_k))$$
$$\leq \frac{(n-1)(n-2)}{2} + V. \quad (9)$$

□

Unfortunately, existing results on network flow do not seem to translate into a polynomial time algorithm for optimizing Objective 20. The second objective, minimizing the time that its last goal is reached, provides a lower bound on the time that is required to reach all goals. Solutions optimizing this objective are useful in providing worst servicing time estimate or guarantee. Solutions to the quickest transshipment problem [19] yield optimal solutions to this objective. However, we can avoid using submodular function optimization routines if we have a polynomial bound on $T$, which is provided in the following corollary of Theorem 17.

**Corollary 23** *There exists an optimal solution for Objective 21 in a time-expanded network with $T = n + \ell - 1$.*

To see that Corollary 23 is true, note that $T = n + \ell - 1$ is sufficient for finding a feasible solution, which must have completion time as large as that of a solution to Objective 21. With the bound on $T$, running $\log T$ rounds (via binary search) of maximum flow over time-expanded network with different time horizon then gives us an optimal solution to Objective 21. The running time is then bounded by $O(\text{MF}(n, V^2, VE) \log V)$, which is strongly polynomial. In particular, with Ford-Fulkerson, the running time becomes $O(nVE \log V)$.

After time optimality, another very useful solution property to optimize is the total distance traveled by the agents, i.e., spatial optimality:

**Objective 24** *Minimizing the total distance traveled by the agents on G.*

Because we work with a CUG, if an agent $a_i$ actually moves along path $p_i$ between time steps $t$ and $t+1$, $p_i(t)$ must be different from $p_i(t+1)$. These correspond to the black edges in the time-expanded network (see Fig. 3(b)). Thus, to optimize this objective, we can find the shortest total distance traveled by all agents via setting the cost of the holdover edge (green edges in Fig. 3(b)) to zero and then running minimum cost maximum flow algorithm over the time-expanded network. The method is again strongly polynomial, with complexity $O(V^2 E \log V)$, due to the following corollary.

**Corollary 25** *There exists an optimal solution for Objective 24 in a time-expanded network with $T = n + \ell - 1$.*

PROOF. Using the scheduling method for establishing the sufficiency condition of Theorem 17 on a path set $Q$ satisfying Properties 8-16 yields the bound. □

## B. Pareto Optimality Between the Objectives

From the discussion in the previous subsection, we observe that each of the three objectives is of practical importance. At this point, one might be tempted to seek solutions that optimize multiples of these objectives simultaneously. We show that this is not possible for each pair of these objectives. In the following theorem, we say that two objectives are *compatible* if and only if they can be optimized simultaneously. Otherwise, we say the objectives are *incompatible*.

**Theorem 26** *Over the feasible solutions to Problem 3, Objectives 20-24 are pairwise incompatible.*

PROOF.
For each pair of objectives from the three objectives, we provide an instance of Problem 3, with which we demonstrate the existence of Pareto optimal solutions.

First we look at Objectives 20 and 21. For this pair we use $G$ from Fig. 8(a) and for the rest of this proof, the red vertices are the initial locations and the blue vertices the goal locations. The optimal solution for Objective 20 is for the agents to take the solid paths connecting the sources and sinks, which has a value of $\frac{3+1+1+1}{4} = \frac{3}{2}$. These paths yield a value of 3 for Objective 21 since the longest path has a length 3. We may write these two values as a vector $(\frac{3}{2}, 3)$. If we optimize for Objective 21, then the dashed paths give a value 2 and these paths yield an average time of $\frac{2+2+2+2}{4} = 2$. In vector form, this is $(2,2)$. Hence, Objectives 20 and 21 are incompatible.

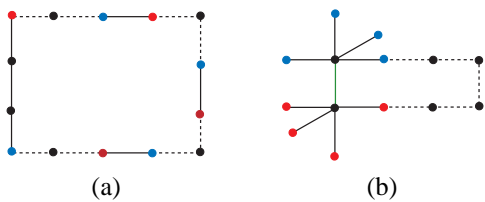

Fig. 8. Flow networks used in the proof of Theorem 26.

To show that Objectives 20 and 24 are incompatible, we use the graph from Fig. 8(b), which can be embeddable in a three dimensional grid (a 2D version can be constructed as well, but that will require more vertices and edges). For Objective 20, the optimal solution is letting the agent on the rightmost red vertex take the path given by the dashed edges and letting the rest of the agents move through the green edge successively, which will in turn require some agents to wait a little. This gives an average time of $\frac{3+4+5+5}{4} = 6.25$. These paths yield a total travel distance of $3+3+3+5 = 14$. If we optimize over Objective 24, then all agents should go through the green edge. This gives an average time of $\frac{3+4+5+6}{4} = 6.5$ and a total travel distance of $3+3+3+3 = 12$. In vector form, we may write these results as $(6.25, 14)$ and $(6.5, 12)$. Clearly, Objectives 20 and 24 are incompatible.

Finally, for Objectives 21 and 24, we may reuse the graph from Fig. 8(a). It is not hard to check that the vectors we obtain for this case should be $(2,8)$ and $(3,6)$, which shows that the objectives are incompatible. □

## VI. PERMUTATION INVARIANT MULTI-AGENT PATH PLANNING PROBLEM ON CUGS WITH GOAL REPLACEMENT

After a thorough discussion of Problem 3, we shift our attention to Problem 4, which allows multiple agents to reach the same goal. An algorithm for finding a feasible solution to this problem is a trivial modification to the algorithm for solving Problem 3: During the construction of the time-expanded network $\mathcal{N}'$, instead of letting the edges $(v', v'')$'s for a goal vertex $v$ have unit capacity, we let them have infinite capacities, which allows the reuse of a goal vertex of $G$ at multiple time steps. An identical version of Corollary 7 can be stated and proved with the same proof. Similarly, Theorem 17 continues to apply. That is, a feasible solution to Problem 4 can be found in strongly polynomial time.

Given feasible solutions to Problem 4, we may again ask optimality questions about these solutions. Objectives 20-24 from Section V still make sense here. Since Problem 3 and 4 are quite similar, it is not surprising (as readers can easily verify) that the same techniques used to optimize these objectives for Problem 3 still apply. For Problem 4, however, another objective on time optimality is also useful in practice.

**Objective 27** *Maximizing the number of agents arriving at each time step $t = 0, 1, \ldots, T$, with smaller $t$'s having higher priorities.*

In the domain of network flow, Objective 27 is often called *earliest arrival flows*. It is applicable to time critical scenarios, such as evacuations due to emergencies (fire, diseases, etc.), in which the environment may become hazardous at any given time and render further evacuation efforts impossible. We did not include this objective in earlier discussions on optimality since the objective is locally greedy and an optimal solution may not be a feasible solution to Problem 3. One simple example is illustrated in Fig. 9(a). Because of the local greedy selection, $s_2^+$ gets paired up with $s_1^-$, which makes goal $s_1^-$ unreachable under the formulation of Problem 3. Also, in case where all goals are reached, Objective 27 may cause later goals to be reached in arbitrarily long time and after traveling arbitrary far. Fig. 9(b) provides an example. Here, local greedy selection forces $s_{i+1}^+$ to be paired up with $s_i^-$ for $1 \le i \le n-1$, which in turn forces $s_1^+$ to be paired up with $s_n^-$. Thus, one agent must go through the upper (dashed) path, which can be made arbitrarily long.

However, when goal replacement is allowed, since each goal can be used arbitrarily many times, problem instances such as these from Fig. 9 are no longer problematic. This makes Objective 27 a good optimization criterion. The objective can be optimized by running minimum cost max flow over the time-expanded network. More interestingly, with goal replacement allowed, Objectives 20, 21, and 27 can be optimized simultaneously [20]. In particular, optimizing Objective 27 guarantees that the other two objectives are also optimized. We may use any optimization method for Objective 27 to optimize all three objectives. The shortest total distance objective, Objective 24, however, cannot be achieved with the other three; the proof of Theorem 26 showing that Objective 20 and 24 are incompatible still holds.

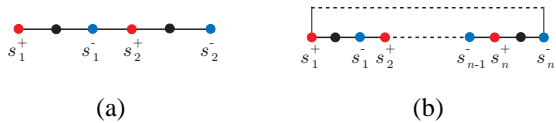

(a)                           (b)

Fig. 9. a) An instance of Problem 3 for which an optimal solution to Objective 27 cannot get to both goals. b) An instance of Problem 3 for which an optimal solution to Objective 27 requires arbitrary long time and distance to reach all goals.

## VII. Conclusion, Future Work, and Open Problems

In this paper, we established the close link between two classes of problems: Multi-agent path planning on CUGs and network flow. Focusing on the permutation invariant versions of the multi-agent path planning problem, we proved a tight bound on the number of time steps necessary and sufficient for a feasible path set to exist in the time-expanded network, enabling efficient algorithmic solutions to these problems. We then explored optimality issues, demonstrating that the time-expansion bound generally carry over to yield strongly polynomial algorithms for optimizing two practical objective functions. Interestingly, each pair of these objectives cannot be optimized simultaneously.

Given our study, an immediate question or criticism is the applicability of the results to problems beyond CUGs. After all, real agents, whether robots or people, do not always live on a discrete graph. To answer this question, we have research under way that explores the idea of overlaying the CUGs on the actual workspace. That is, we may first create a roadmap over the workspace that captures the connectivity and then discretize the roadmap over which the statement of Observation 2 continues to hold (as long as the edges are close to unit length the angle between two edges is obtuse, similar version of Observation 2 can be stated) [35]. A basic solution (Fig. 10(a)) may be to put a grid on the roadmap and delete vertices inside or close to obstacles. To overcome the issue of the inherited Manhattan metric of grids, we may adapt the grid to align with the geodesics of the environment. For example, for a two dimensional workspace with polygonal obstacles, we can arrange the grid edges to follow edges of the visibility graph [29] of the environment when possible. When clearance is tight, we may start with a maximum clearance roadmap [35] and add the vertices carefully (see Fig. 10(b)). Note that because the workspace is often two dimensional, these preparations can be computed relatively efficiently.

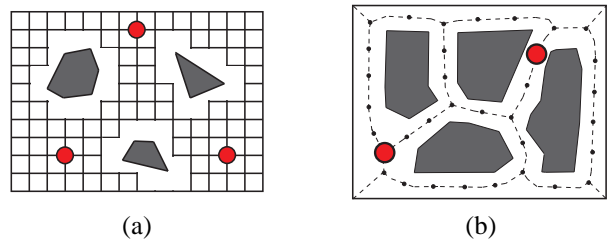

(a)                           (b)

Fig. 10. a) Overlaying grid on a workspace with obstacles. b) Adapting a roadmap to obtain a graph that can be used with our multi-agent path planning algorithms.

Many interesting open problems remain. Although finding a distance optimal solution to Problem 1 using a time-expanded network is impractical due to its intrinsic hardness, the network flow approach might still produce efficient methods that yield basic feasible solutions since the time-expanded network has a forward only structure. In addition, approximation algorithms on integer multi-commodity flow could lead to better heuristics for optimal solution search. Along this line, we only touched the most essential results in the field of network flow, which are but the tips of an iceberg. It would not be surprising that results from the vast amount of network flow literature could be readily carried over to tackle path planning problems, as we proposed in this paper or in some other forms. As an example, for Problem 3, since Objectives 20-24 are all of practical concerns but incompatible, it is desirable to seek solutions that provide performance guarantees on each of these objectives. Network flow methods, closely relate to linear programming, appear to be promising tools for such parametric optimization tasks.

**Acknowledgments** This work was supported in part by NSF grants 0904501 (IIS Robotics) and 1035345 (Cyberphysical Systems), DARPA SToMP grant HR0011-05-1-0008, and MURI/ONR grant N00014-09-1-1052.